\documentstyle[prl,aps,floats,epsf,twocolumn]{revtex}

\begin{document}
\draft
\flushbottom
\twocolumn[\hsize\textwidth\columnwidth\hsize\csname @twocolumnfalse\endcsname

\title{Energetics of the oxidation and opening of a carbon nanotube}

\author{M. S. C. Mazzoni,$^*$ H. Chacham,$^*$
P. Ordej\'on,$^{\dagger}$ D. S\'anchez-Portal,$^{\ddagger}$
J. M. Soler,$^{\ddagger}$ and E. Artacho$^{\ddagger}$}

\address{
$^*$Departamento de F\'\i sica, ICEx, Universidade Federal de
Minas Gerais. C.P. 702, 30123-970 Belo Horizonte, MG, Brazil \\
$^{\dagger}$Departamento de F\'\i sica, Universidad de Oviedo, 
33007 Oviedo, Spain \\
$^{\ddagger}$Departamento de F\'\i sica de la Materia Condensada, C-III,
Universidad Aut\'onoma de Madrid, 28049 Madrid, Spain}

\date{27 May 1999}
\maketitle

\begin{abstract}
We apply first principles calculations to study the opening of
single-wall carbon nanotubes (SWNT's) by oxidation.
We show that an oxygen rim can stabilize the edge of the open tube.
The sublimation of CO$_2$ molecules from the rim with the subsequent
closing of the tube changes from endothermic to exothermic as the tube
radius increases, within the range of experimental feasible radii.
We also obtain the energies for opening the tube at the cap and 
at the wall, the latter being significantly less favorable. 
\end{abstract}


]

The possibility of filling carbon nanotubes\cite{1} with a variety of  
substances has fascinated the materials science community since they
were first observed \cite{2} and synthesized in promising
quantities \cite{3}, but the fact that the tube tips are closed
represented a severe limitation for it.
Ajayan and Iijima \cite{4} observed the cap removal and
filling of the nanotubes when they were heated in air,
suggesting a chemical reaction between oxygen and the tube ends. 
This was confirmed by Ajayan et al \cite{5} who showed that the 
tubes were open exclusively at the caps.  
Simultaneously, Tsang, Harris and Green \cite{6} described 
the thinning and opening of nanotubes by carbon dioxide. 
They also observed corrosion at the cap region, the tube walls 
remaining intact.
Ugarte et al \cite{7} described a method of filling  nanotubes
in which they were first oxidized in air to remove the caps. 
In this process, $60\%$ of the tubes were open.
The above mentioned experiments used multi-walled nanotubes.
In a recent experiment, however, Dillon {\it et al.} \cite{8} 
observed the physisorption of H$_2$ in bundles of small-diameter SWNT's.
They suggested that these tubes were open by          
oxidation in the same way as multi-walled tubes. 
Individual open-ended SWNT's were also obtained experimentally 
\cite{9} by treating SWNT rope material with a mixture of acids.

The filling and opening of nanotubes have also been subjects of theoretical 
studies. Pederson and Broughton \cite{10} predicted the tendency of the  
nanotubes to attract molecules to their interior. More recently,
Miyamoto et al \cite{11} reported first-principles calculations 
that showed the ionic cohesion between carbon nanotubes and potassium 
atoms linearly arranged inside them.
The reactivity of the open end of SWNT's was studied by 
Charlier et al \cite{12}. They showed that the unsaturated tip of small
diameter SWNT's closes spontaneously at typical growth 
temperatures. However, first principles calculations also indicate  that
these tubes may stay open during growth in the presence of a catalyst, 
such as Ni \cite{13}.

The detailed description of the burning and opening processes, and of 
their kinetics, is a formidable problem beyond the scope of state-of-art 
ab initio techniques. In the present work, 
we apply first principles calculations to address two specific,
but relevant questions on the energetics of these processes:
($i$) why does O$_2$ attack preferentially the tube caps?
and ($ii$) how and why do the tubes remain open (if at all) after 
the O$_2$ supply has been cut? 
To address the first question, we compare the opening of oxidized holes 
at the cap of the tube with their chemical equivalents at the wall,
finding that the cap oxidation is much more favorable, due to the 
release of its curvature strain. 
On the second question, we find that the presence of an oxygen rim 
at the edge of the open tube results in a very stable configuration,
even in the absence of O$_2$.
Smaller holes with favorable energetics allow us to obtain an estimate of
the activation barrier of the closing process. 

Our calculations were performed with the SIESTA program \cite{14,15,16},
using density-functional theory \cite{17}, 
within the generalized gradient approximation (GGA) 
for exchange-correlation \cite{18}, 
and norm-conserving pseudopotentials \cite{19,20}. 
The basis set is a linear combination of pseudoatomic orbitals 
\cite{21,foot_r_c}. 
We always used a split-valence double-$\zeta$ (DZ) basis
\cite{foot_split}, and convergence tests were performed by adding 
polarization functions \cite{22}. 
This resulted in changes in formation energies of only a few meV/atom.
All the geometries were optimized, 
with remanent forces of less than 0.1 eV/\AA.
Test calculations were performed on graphite, nanotubes, and fullerenes.
The calculated  nearest-neighbor distances of C$_{60}$ (1.42 and
1.47 \AA) and graphite (1.42 \AA) reproduce the experimental values
(1.40, 1.46, and 1.42 \AA, respectively)\cite {1} within 2\%. Also, the
energy per atom of C$_{60}$ relative to that of graphite (0.38 eV/atom) agrees 
very well  with the experimental result (0.392 - 0.433 eV/atom)\cite {24}. 
Our calculations were performed for (4,4), (6,6), and (8,8) armchair SWNT's, 
with diameters 5.6, 8.3, and 11.1 \AA,
containing up to 166 carbon atoms.
However, for the sake of clarity in the exposition, we will refer to the
specific (6,6) case throughout the paper, leaving the analysis
of the size effects to the end.

\begin{figure}[t!]
\narrowtext
\vspace{-3.5cm}
\centering
\epsfxsize=1.8\linewidth
\epsffile{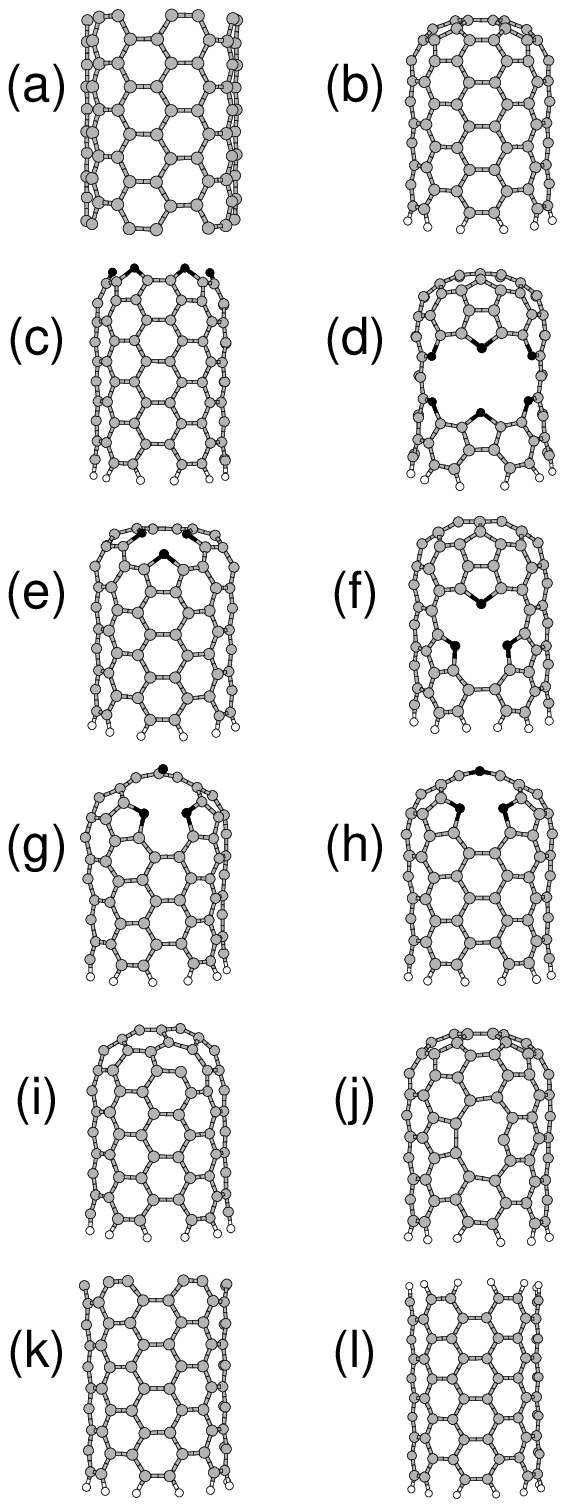}
\vspace{-6 cm}
\caption[struc]{
Structures used in the calculations for the (6,6) carbon nanotube. 
(a) Infinite tube (periodic boundary conditions); 
(b) closed tube, with 120 carbon atoms and 12 hydrogen atoms saturating
the bulk-like end of the tube; (c) oxidized open tube; (d) oxidized
opening equivalent to (c) but at the wall instead of the cap;
(e) oxidized hole obtained by removing an hexagon at the cap; (f) same as (e)
at the wall; (g) oxidized hole obtained by removing a pentagon; (h)
same as (g) but removing an additional C atom; (i) a C vacancy at the cap;
(j) vacancy at the wall; (k) unsaturated open tube; (l) H-saturated
open tube on both sides (used to characterize the saturation energy).
The free H$_2$, O$_2$, H$_2$O, and CO$_2$ molecules were also calculated.
The back atoms are not shown in the figure.}
\label{fig:1}
\end{figure}

In order to understand why O$_2$ produces a selective oxidation
of the tube caps, rather than a generalized burning,
we compare the energies of similar holes at the tube's cap and wall.
Figure 1 shows these structures for the (6,6) tube.
The closed tube, hereafter denoted by C$_{120}$, is shown in Fig.~1(b).  
The other end of the tubes is bulk-like and saturated by hydrogen atoms. 
This termination is used in all the structures, except the infinite tubes 
used to calculate the carbon chemical potential $\mu_C$, for which we 
used periodic boundary conditions [Fig.~1(a)].
The hole geometries, shown in Fig. 1(c)-(j), were chosen as plausible
low energy strutures for specific hole sizes and not necessarily to 
simulate the actual oxidation process. These hole structures will 
provide estimates of the energy released, for a given
hole size, during the tube opening process.
Figure 2 shows the potential energy change during oxidation as a function of 
the hole size, measured by the number of carbon atoms lost by the tube in the 
process. These atoms are removed from the tube according to the reaction
C$+$O$_2\rightarrow$ CO$_2$, while extra oxygen is used to
saturate the resulting holes. The initial structure is Fig.~1(b)
and the oxidized structures are Fig.~1(c)-(j). The general trend is
linear with the number of C atoms missing in the hole, as it should, but
the energy released by oxidizing the cap is larger than for the wall, as
shown in Fig.~2(b) \cite{assum}. 
As we will see, the origin of the difference is the extra elastic energy
accumulated at the cap.
It is remarkable that the wall-cap difference for removing the first 
carbon atom (2.4 eV), is already very close to that for a full 
hexagonal hole.
Thus, the cap's strain energy is substantially released from the very
beginning of the oxidation.

Addressing the stability of the open 
tube after oxidation, Fig~1(c) shows the most efficient chemical 
saturation of the open tube with oxygen. In this oxidized open tube, 
denoted by C$_{120}$O$_6$, every O atom saturates two carbon ``dangling 
bonds", giving a slightly stretched C-O bond distance of 1.43 \AA~and
a C-O-C angle of 105$^{\rm o}$. The bond stretching produces an
inward relaxation (18$^{\rm o}$) of the rim structure 
as a consequence of the pulling of the O atoms, diminishing the tube perimeter
at the rim. The structure is very stable: the sublimation of 
oxygen atoms in the form of CO$_2$, resulting in 
an unsaturated open tube [Fig.~1(k)], is found to be energetically unfavorable 
by 17.8 eV. This highly endothermic reaction indicates that, if open tubes
are present, they will  have an oxygen rim at the edge even at typical
oxidation temperatures ($\sim$ 1000K) \cite{25}. 

\begin{figure}[t!]
\narrowtext
\vspace{-0.2cm}
\centering
\epsfxsize=0.8\linewidth
\epsffile{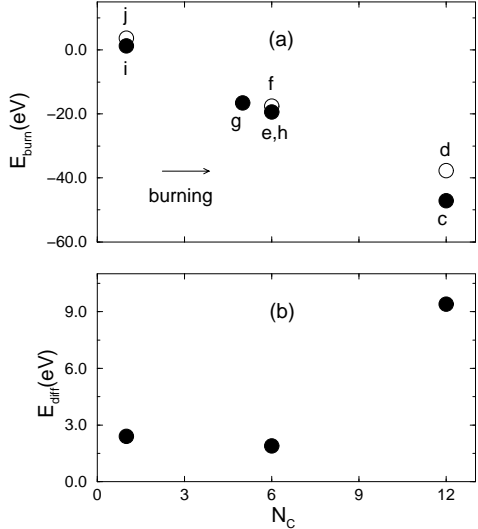}
\vspace{-1.6 cm}
\caption[burn]{
(a) Potential energy change, E$_{burn}$, during the burning process
of the original closed tube, leaving behind the holes shown in 
Fig.~1. E$_{burn}$ is plotted versus the number of C atoms missing in the final
hole. Filled symbols are for holes at the cap of the tube, open symbols
for holes at the wall. The labels refer to the structures of Fig. 1.
(b) Energy difference between burning the tube at
the wall or at the cap: $E_{diff} = E_{burn}^{wall}-E_{burn}^{cap}$ .}
\label{fig:2}
\end{figure}

Starting with this oxidized open tube, we consider a reaction in which 
the tube closes and CO$_2$ molecules evaporate: 
\begin{equation}
{\rm oxidized~open~tube}  \rightarrow {\rm closed~tube} + 3{\rm CO}_2.
\label{eq1}
\end{equation}
The energy of this reaction is  
$ \Delta E = (E_{{\rm C}_{120}} + 3E_{{\rm CO}_2}) -
             (E_{{\rm C_{120}O}_6} + 3\mu _{\rm C}) $
The result is $\Delta E =  0.2$ eV with the DZ basis, and
$\Delta E= 0.3$ eV including polarization orbitals.
Thus, the oxidized open tube is slightly favorable energetically over the
closed one for this tube diameter.  
However, it has to be stressed that 
the system composed by the closed tube and the carbon dioxide gas 
has larger entropy and will thus predominate in thermodynamic equilibrium 
at a high enough temperature.

In addition to its genuine stability, the oxidized open tube can
be further stabilized by the energy barrier of 
the closing process, which requires substantial rebonding. 
The actual rebonding is extremely hard to guess 
or simulate. We search for an estimate of the energy barrier
as follows: first we consider a ``closing coordinate" that the system
must necessarily follow during the closing process. We simply use the hole 
size, defined as before as the number of carbon atoms missing from the closed
tube. Then, for particular values of this coordinate, we search for
low energy structures using basic chemical principles: to break the minimum
number of the most strained C-C bonds, and to saturate them
with oxygen. The chosen structures are those in Fig. 1(e), (g), and
(h), obtained by removing an hexagon, a pentagon, and a pentagon
plus an extra C atom, respectively, and introducing three O atoms
saturating the holes. 
We shall mention that other small-hole structures or other 
closure mechanisms (with catalysts, for instance) with even lower energies
can possibly exist, our construction giving just a plausible estimate
of a low energy barrier.
Again, we considered reactions in which
the small holes close and CO$_2$ molecules evaporate, and we evaluated
the energy difference between the initial and final states. The results
are summarized in Fig.~3, which shows (solid line) these hole energies
as a function of the number of carbon atoms missing in the hole. 
We can see in the figure that for closing the tube just 0.2 eV
are needed, but a barrier of at least 4.2 eV has to be overcome first.
Remarkably, the hole formation energies for the three small holes
considered here are degenerate within 0.1 eV.
 As for the equivalent wall holes, displayed in Fig.~1(d) and (f),
Fig.~3 shows (dashed line) that the tube closing is clearly exothermic
(by 9.2 eV) and that no energy barrier is found.
                                               
We can understand the origin of the distinct cap and wall behaviors
by defining a cap energy $E_{cap}$, as the energy of the capped tube
minus that of the same number of atoms in the bulk of the tube.
Since chemical bonds are similarly saturated at the cap and at the wall,
the origin of $E_{cap}$ must be the larger elastic energy of the cap.
It amounts to 9.0~eV for the (6,6) tube, which is comparable to the 9.2~eV
energy obtained for the opening of the large hole
at the tube wall. 

\begin{figure}[b!]
\narrowtext
\vspace{-3cm}
\centering
\epsfxsize=0.85\linewidth
\epsffile{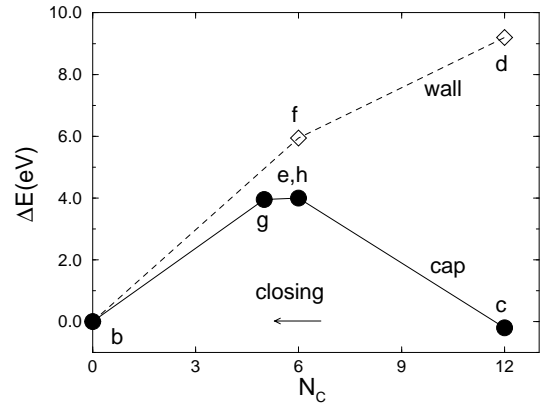}
\vspace{-1.5cm}
\caption[closing]{
Energy differences between initial and final states of the closing     
reaction: $tube~with~hole \rightarrow closed~tube + CO_2$.  The energies
are shown as a function of the number of carbon atoms missing in the 
hole. Solid and dashed lines 
are used for holes at the cap and wall, respectively. The labels refer
to the structures of Fig. 1.}
\label{fig:3}
\end{figure}

Since the release of the strain energy, which plays a crucial role
in the opening process, is related to the tube diameter,           
narrower tubes should be more easily open due to 
their larger cap curvature.
Our calculations on the (4,4) and (8,8) armchair tubes confirm this trend:
with $CO_2$ molecules used as an oxygen atom reservoir, as in equation (1),  
the closing energy amounts to 1.2 eV for the (4,4) tube, 
0.2 eV for the (6,6) tube (as we have already shown), and -0.3 eV for 
the (8,8) tube, allowing us to estimate a value of $\sim -0.6$ eV for the 
(10,10) tube. Therefore, within the range of experimental 
feasible radii \cite{26} the CO$_2$ sublimation changes from endothermic to 
exothermic, always within tenths of eV.

The presence of an oxygen rim may have a strong influence on
the ability to fill the tubes. 
In some cases, it may be desirable to replace it by 
a possibly more inert hydrogen rim.
Therefore, we have studied the energetics of reducing the 
oxidized open tube with H$_2$.
Simply replacing O with H, followed by O$_2$ evaporation,
is slightly endothermic (by 0.3 eV for the (6,6) tube).
However, the reaction becomes strongly exothermic, 
by 9.1 and 9.5 eV, when we consider CO$_2$ and H$_2$O as byproducts,
respectively.
Thus, we expect that the reduction of the oxygen rim should be 
quite possible.
We shall mention that the saturation of the tube ends with other
molecular terminations rather than O or H could result in even lower
formation energies.

In conclusion, we have studied the energetics of opening SWNT's
by oxidation. 
We show that an oxygen rim can stabilize the edge of the open tube.
The sublimation of CO$_2$ molecules from the rim with the subsequent
closing of the tube changes from endothermic to exothermic as the tube
radius increases, within the range of experimental feasible radii.
Our results also show that the opening reaction occurs      
preferably at the tube cap, due to the release of its strain energy.

\acknowledgements
We acknowledge support from the Brazilian agencies CNPq, FAPEMIG,
CAPES, and PRONEX-MCT,
and from Spanish DGES grant PB95-0202. P.O. is the recipient of a 
Motorola PCRL Sponsored Research Project.
We also thank R. W. Nunes for a critical reading of the manuscript.


\begin{references}
\bibitem{14}
The SIESTA program is described in Ref. \cite{16}. 
It builds the DFT selfconsistent
Hamiltonian in order-$N$ operations. The Hamiltonian problem
can be solved either with diagonalization (order-$N^3$) or
with order-$N$ algorithms according to convenience. In this
study both options were used.
\bibitem{foot_r_c} 
The range of the orbitals is defined in a balanced way 
by a common orbital-confinement energy of 0.14 eV \cite{22}. 
The resulting radii  are
4.52 and 5.52 for the $s$ and $p$ orbitals of C, respectively, 5.20 for
the $s$ orbital of H, and 3.56 and 4.35 for $s$ and $p$ orbitals of O, 
all of them in atomic units.
\bibitem{foot_split} The split-off function reproduces the tail of the 
original orbital beyond some radius, matching to a smooth $r^l (a - b r^2)$
towards the nucleus. The matching radius is chosen so that the
split orbital carries 15 \% of the norm of the original one \cite{22}.
\bibitem{assum}
Note that the numbers in Fig.~2(a) depend on the proposed oxidation process, 
and that other possibilities, like involving C and O 
interstitials are plausible, and would affect the results. The origin of
the ambiguity is the presence of three references (bulk C, CO$_2$, and O$_2$)
for defining the chemical potential of just two species, C and O. The process 
proposed involves no extra C atoms. However, the energy differences shown 
in Fig.~2(b) are independent of these assumptions. There is no ambiguity 
in the case of the closing process, since, in the absence of O$_2$,
both chemical potentials are well defined.
\bibitem{25}
At  growth temperatures ($\sim$ 3000K) , molecular dynamics simulations
have shown that narrow SWNT's close spontaneously \cite {12}.
\end{references}
\end{document}